\documentclass[11pt]{elsarticle}
\usepackage[margin=1.2in]{geometry}
\usepackage{gensymb}
\usepackage{graphicx}
\usepackage{amsmath}
\usepackage{latexsym}
\usepackage{amsfonts}
\usepackage{mathtools}
\usepackage{tcolorbox}
\usepackage{array}
\usepackage{subcaption}
\usepackage{changepage}
\usepackage{dirtytalk}
\usepackage{natbib}
\usepackage[makeroom]{cancel}
\usepackage{amsthm}
\usepackage{epstopdf}
\usepackage{amssymb}
\usepackage{a4wide}
\usepackage{pst-all}
\usepackage{float}
\usepackage{latexsym}
\usepackage{graphicx}
\usepackage{psfrag}
\usepackage{verbatim}
\usepackage{setspace}
\usepackage{xcolor,colortbl}
\usepackage{relsize}
\usepackage{float}
\usepackage{pifont}
\usepackage[font={small}]{caption}
\usepackage[labelsep=space]{caption}
\usepackage{upgreek}

\DeclareCaptionFormat{myformat}{\fontsize{9}{10}\selectfont#1#2#3}
\makeatletter
\renewcommand{\fnum@figure}{\textbf{Fig. \thefigure.}}
\makeatother

\makeatletter
\renewcommand{\fnum@table}{\textbf{Tab. \thetable.}}
\makeatother

\renewcommand*\thetable{\Roman{table}}

\newcommand\fesum{\mathop{%
  \ooalign{%
    \hfil$\displaystyle\sum$\cr
    \hfil$\textstyle\sum$\cr}}}

\newcommand{\ds}{\displaystyle}
\newcommand{\bm}[1]{\text{\boldmath $#1$}}
\newcommand{\bs}[1]{\textsf{\textbf{#1}}}
\newcommand{\norm}[1]{\left\lVert#1\right\rVert}
\newcommand{\pfrac}[2]{\frac{\partial #1}{\partial #2}}
\newcommand{\grad}{\bm{\nabla}}

\makeatletter
\def\ps@pprintTitle{%
 \let\@oddhead\@empty
 \let\@evenhead\@empty
 \def\@oddfoot{}%
 \let\@evenfoot\@oddfoot}
\makeatother

\begin{document}

\begin{frontmatter}
	\title{Shape optimization of pneumatic soft actuators}
	\author{Anna Dalklint\fnref{label1,label2}\corref{cor1}}
	\author{Vilmer Dahlberg\fnref{label2}}
    \author{Katia Bertoldi\fnref{label1}}
	\address[label1]{Harvard John A. Paulson School of Engineering and Applied Sciences, Harvard University, Cambridge, MA, USA \fnref{label1}}
	\address[label2]{Division of Solid Mechanics, Lund University, Box 118, SE-22100 Lund, Sweden \fnref{label2}}
    \cortext[cor1]{Corresponding author. E-mail address: anna\_dalklint@seas.harvard.edu}

\begin{abstract}
Soft actuators, characterized by their compliance and flexibility, have tremendous potential for diverse applications, ranging from medical devices to submarine operations. However, significant challenges remain in the design of these actuators, specifically in maintaining precise control over their mechanical behavior and motion. To date, heuristic methods have been commonly used to design soft actuators, which are potentially incapable of producing designs that achieve specific target behaviors. We propose a gradient-based inverse design framework to synthesize three dimensional soft actuators with tailored mechanical responses. Our design framework utilizes gradient information that captures the inherent geometrical and material nonlinearities of the soft actuator to morph its shape. We exemplify the capabilities of the proposed framework by designing soft actuators with bespoke deformation patterns, making use of sophisticated deformation mechanisms to realize the target behavior. The capabilities of the proposed framework are validated via experimental testing of cast designs, which confirms a strong correlation between measurements and numerical simulations. 


\end{abstract}


\end{frontmatter}


\section{Introduction}
Soft actuators are compliant, flexible structures that generate motion or force through the deformation of soft materials such as elastomers, hydrogels, or shape-memory polymers. These actuators have demonstrated significant potential across a broad range of applications, including assisting surgeons during minimally invasive procedures (\citet{runciman2019soft}), manipulating delicate objects (\citet{shintake2018soft}) and navigating hazardous or constrained environments (\citet{hawkes2017soft}, \citet{li2023bioinspired}). Various actuation mechanisms have been developed for these systems, including pneumatic and hydraulic pressure (\citet{ilievski2011soft}, \citet{mosadegh2014pneumatic}, \citet{gorissen2020inflatable}), electric fields (\citet{anderson2012multi}, \citet{shian2015dielectric}), chemical reactions (\citet{wehner2016integrated}) and magnetic fields (\citet{kim2022magnetic}). Among these approaches, fluidic actuation has emerged as one of the most widely adopted methods due to its structural simplicity, ability to achieve large deformations and straightforward fabrication processes.

Traditionally, the design of soft actuators is based on heuristics, relying on engineering intuition and parameter space exploration (\citet{mosadegh2014pneumatic}, \citet{shepherd2011multigait}). Naturally, such methods are potentially incapable of producing designs that achieve specific target behaviors. In this regard, design methods based on gradient-based optimization are much more efficient, whereby the optimized material distribution is provided under given objective and constraint functions. 
Gradient-based optimization in the form of shape or topology optimization has proven its use for designing pneumatic soft actuators using implementations based on finite strain hyperelasticity (\citet{mehta2025topology}, \citet{dalklint2024simultaneous} and \citet{caasenbrood2020computational}) and linear elasticity (\citet{de2020topology}, \citet{lu2022optimal} and \citet{kumar2025soft}). However, the aforementioned works are limited to shape morphing in two dimensions, which naturally limits the methods capabilities of finding truly novel designs with enhanced capabilities. Exception to this rule is the work of \citet{kobayashi2024computational}, who integrates gradient-based topology optimization with a material point method to design locomotive soft robots, and \citet{chen2023morphological} who proposes a shape optimization scheme based on B-splines to design soft actuators with desired deformation behavior. 

Shape optimization is an excellent candidate for the design of pneumatic soft actuators, since the pressurized internal cavities of the actuators naturally lend themselves to shape morphing. Indeed, topology optimization should in theory be able to explore a more vast design space, but it is not entirely clear how to simultaneously control each independent air chamber's actuation pressure. Also, the gas or fluid pressure must somehow be transferred to the solid body, which is nontrivial when the design boundaries are a priori unknown, and entails additional modeling of e.g. incompressible fluid regions (\citet{sigmund2007topology}, \citet{bruggi2009alternative}) or porohyperelasticity (\citet{mehta2025topology, mehta2026topology}). A shape optimization framework has previously been developed to design three dimensional soft actuators by \citet{chen2023morphological}. Their approach is based on tracking the evolving surfaces of the hyperelastic body during optimization using B-spline surfaces, and they regulate the surface quality by imposing geometric penalty constraints. Their computational workflow is not unified, but subdivided such that different pieces of software handle geometry, finite element modeling and optimization. This comes with a high computational cost, wherefore they must rely on solving a sequence of subproblems that are constructed within a trust region in which the cost and constraint functions are approximated by their first-order Taylor polynomials.

The present work proposes a unified computational framework capable of designing shape optimized three dimensional soft actuators, taking into account both material and geometrical nonlinearities. The computational framework is based on a nearly incompressible hyperelastic finite deformation (FE) implementation on unstructured meshes that utilizes the Portable and Extendable Toolkit for Scientific Computing (PETSc) for efficiency (\citet{balay2019petsc}). The shape optimization is simple in the sense that it is parameter-free; it directly utilizes the nodal coordinates in the finite element mesh as design variables. In this approach to shape optimization, originally proposed by \citet{scherer2010fictitious}, the mesh quality is regulated via a non-linear PDE filter (\citet{dalklint2023computational}, \citet{dahlberg2025simultaneous}). The design is updated using the gradient-based method: Method of Moving Asymototes (MMA, \citet{svanberg1987method}), and the gradients of the cost and constraint functions are obtained from an adjoint sensitivity analysis. We demonstrate the effectiveness of the proposed framework by designing actuators capable of performing nontrivial deformations upon actuation, such as object grasping, contraction under pressurization and staggered multimodal deformation modes. Finally, we validate the computational framework by fabricating the optimized designs using a casting approach and experimentally characterize their performance. 


\section{Hyperelastic formulation}
The response of the hyperelastic body is captured using a Lagrangian kinematic description in which $\Omega$ is the undeformed configuration with material points $\bm{X}\in\Omega$. A pressure load deforms $\Omega$ into the current configuration $\Omega_c=\bm{\varphi}(\Omega)$, with spatial points $\bm{x} = \bm{\varphi}(\bm{X}) = \bm{X} + \bm{u}(\bm{X})$, where $\bm{\varphi}$ is the smooth deformation and $\bm{u}$ is the corresponding displacement. The local deformation is described by the deformation gradient $\bm{F} = \bm{\nabla} \bm{\varphi} = \bs{1} + \bm{\nabla}\bm{u}$, Jacobian $J(\bm{F}) = \text{det}(\bm{F})$, right Cauchy-Green deformation tensor $\bm{C}(\bm{F}) = \bm{F}^T\bm{F}$ and the Green-Lagrangian strain $\bm{E}(\bm{C}) =\frac{1}{2}\left(\bm{C} - \bs{1}\right)$, where $\bm{1}$ is the second order identity tensor. 

\subsection{Material model}
We cast the soft actuator from an elastomer that is assumed to be isotropic, hyperelastic and nearly incompressible, wherefore it is modeled using the neo-Hookean strain energy density
\begin{equation}
 {W}(\bm{F}) = \Psi(J(\bm{F}), \bm{C}(\bm{F}))=\frac{1}{2}G\left(J^{-2/3}\text{tr}(\bm{C}) -3\right)+ \frac{1}{2}K(J- 1)^2,
\label{c1}
\end{equation}
where, in the limit of infinitesimal strain, $K = \frac{E}{3(1-2v)}$ and $G = \frac{E}{2(1+\nu)}$ denote the bulk and shear modulii, respectively, and $E$ is the Young's modulus. To obtain a nearly incompressible behavior, we let the Poisson's ratio $\nu = 0.49$, such that $J(\bm{X})\approx 1$, $\forall \bm{X}\in\Omega$. To resolve the issue of nonphysical oscillations in pressure when the stress is computed directly from the strain energy function for nearly incompressible materials, we use the mixed formulation proposed by \citet{sussman1987finite}. In this formulation, we introduce an independent pressure field $\tilde{p}$ such that we can define a modified strain energy function
\begin{equation}
\widetilde{W}(\bm{F}, \tilde{p}) = W(\bm{F}) - \frac{1}{2K} \left(p-\tilde{p}\right)^2,
\label{Wmod}
\end{equation}
where the hydrostatic pressure, $p$, is derived from the original constitutive relation, i.e. Eq. \eqref{c1} as
\begin{equation}
p = -\frac{\partial \Psi}{\partial J}=- K(J -1). 
\label{p}
\end{equation} 
For later purposes, we note that the second Piola-Kirchhoff stress tensor is $\widetilde{\bm{S}} = \bm{F}^{-1} \pfrac{\widetilde{W}}{\bm{F}} = 2\frac{\partial \widetilde{W}}{\partial \bm{C}}$. 

\subsection{Equilibrium equation}
The equilibrium configuration of the elastomer is defined by those sufficiently smooth fields $\bm{u}$ and $\tilde{p}$ which, for all admissible virtual fields $\delta \bm{u}$ and $\delta \tilde{p}$, satisfy 
\begin{equation}
\ds\delta\Pi(\bm{u},\tilde{p};\delta\bm{u},\delta\tilde{p}) = \ds\int_\Omega \widetilde{\bm{S}}: \delta \bm{E} \, dV 
\ds+ \int_\Omega \frac{1}{K}\left(p - \tilde{p} \right)\delta \tilde{p} \, dV+\int_{\partial\Omega_{c}^{p}}\hat{p} \bm{n}_c\cdot\delta\bm{u} \, dS =0,
\label{deltaW01}
\end{equation}
where $\delta\bm{E} =\frac{1}{2}\left((\bm{\nabla}\delta\bm{u})^T\bm{F} + \bm{F}^T \bm{\nabla}\delta\bm{u}\right)$ is the virtual Lagrangian strain. In the above, the deformed boundary $\partial\Omega_c$ with the unit normal $\bm{n}_c$, consists of two complementary surfaces, $\partial\Omega_{c}^{p}$ and $\partial\Omega^{u}=\partial\Omega_{c}^{u}$, over which the pressure load $\hat{p}\bm{n}_c$ and the null displacement $\bm{u} = \bm{0}$, are prescribed, respectively. It is emphasized that $\hat{p}$ is the relative pressure compared to the atmospheric pressure.  

\section{Shape optimization}
Our shape optimization is defined similar to the aforementioned hyperelastic problem in the sense that it morphs the initial design $\Omega_o$ into the shape optimized design $\Omega$ via a smooth vector field $\bm{\psi}$, which takes points $\bm{X}_o\in\Omega_o$ to
points $\bm{X}\in\Omega$ via $\bm{X} = \bm{\varphi}_\psi(\bm{X}_o) = \bm{X}_o + \bm{\psi}(\bm{X}_o)$. The field $\bm{\psi}$ is driven by the shape optimization design variable vector field $\bm{d}$ via a PDE filter defined by the potential (\citet{scherer2010fictitious}, \citet{swartz2023yet}, \citet{dahlberg2025simultaneous})
\begin{equation}
\Pi_\psi(\bm{\psi};\bm{d}) = \ds\frac{1}{2}\int_{\Omega_o} W_\psi(\bm{F}_\psi) \, dV + \frac{1}{2\vert\Omega_o\vert}\int_{\Omega_o} \vert\vert\bm{\psi}-\bm{d}\vert\vert^2 \, dV,
\label{PDEpot01}
\end{equation}
where $W_\psi$ is a fictitious strain energy density function. We also introduce the shape field deformation gradient $\bm{F}_\psi = \bm{\nabla} \bm{\varphi}_\psi= \bs{1} + \grad\bm{\psi}$. 
The smooth shape field $\bm{\psi}$ is found by invoking stationarity in Eq. \eqref{PDEpot01}, i.e.
\begin{equation}
\delta\Pi_\psi(\bm{\psi};\bm{d},\delta\bm{\psi}) = \ds\int_{\Omega_o} \bm{S}_\psi : \delta\bm{E}_{\psi} \, dV + \frac{1}{\vert\Omega_o\vert}\int_{\Omega_o} \left(\bm{\psi}-\bm{d}\right) \cdot \delta\bm{\psi} \, dV = 0,
\label{PDEpot02}
\end{equation}
which should be fulfilled for all admissible virtual fields $\delta\bm{\psi}$. We prescribe the null displacement $\bm{\psi} = \bm{0}$ over $\partial\Omega_o^\psi$.

\begin{figure}
    \centering
    \includegraphics[width=0.7\textwidth]{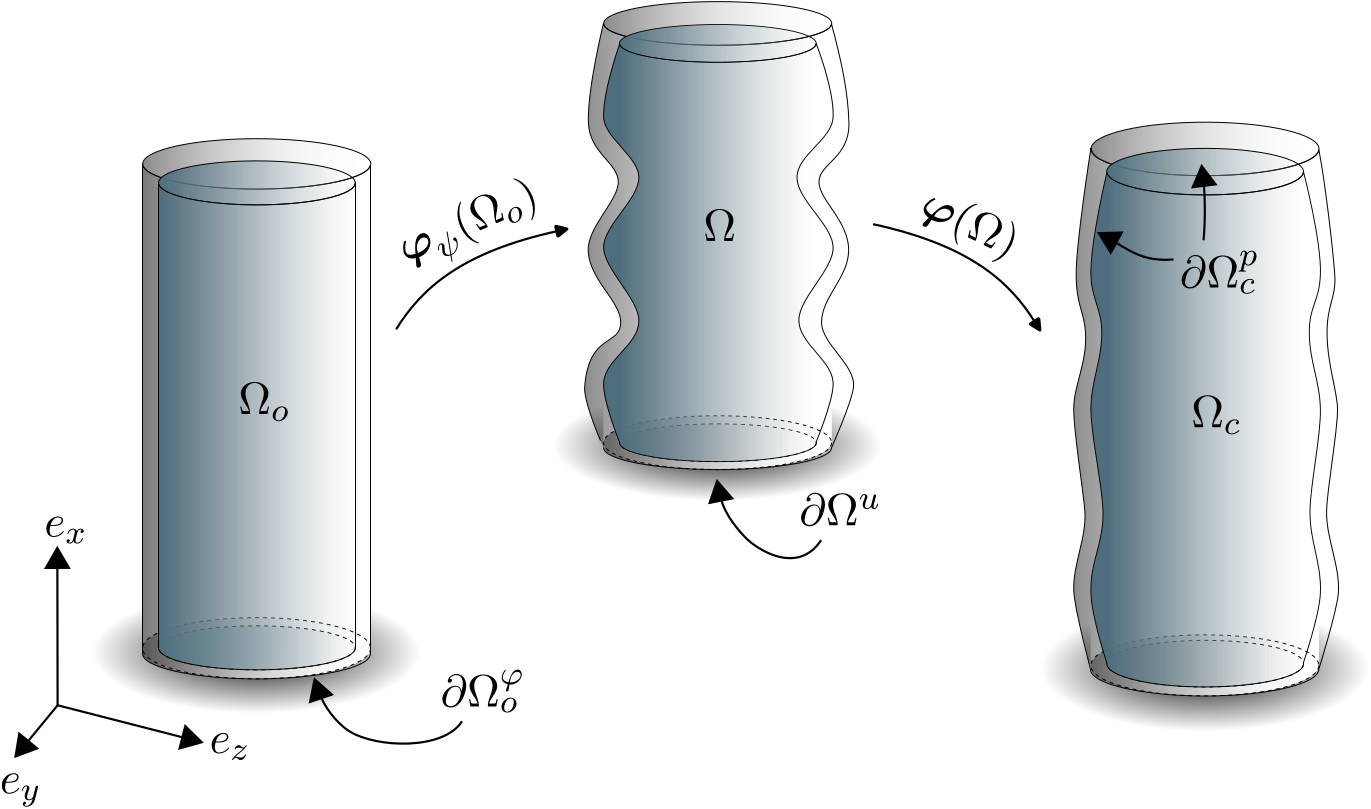}
    \captionof{figure}{\textbf{Mappings}. Illustration of the mappings $\bm{\varphi}_\psi$ and $\bm{\varphi}$.}
    \label{fig:potato}
\end{figure}

The framework for mapping the initial domain $\Omega_o$ to the shape optimized domain $\Omega$ to the deformed domain $\Omega_c$ is illustrated in Fig. \ref{fig:potato}.

\subsection{Fictitious material model}
The fictitious strain energy density function $W_\psi$ should be chosen such that it adequately smooths the shape field $\bm{\psi}$. In the literature, several possible energy density functions have been proposed, see also \citet{swartz2023yet}. In this work, we follow \citet{dahlberg2025simultaneous} and use the neo-Hookean like energy density
\begin{equation}
\begin{array}{ll}
\ds W_\psi(\bm{F}_\psi) &=\ds\Psi_\psi(J_\psi(\bm{F}_\psi), {\bm{C}}_{\psi}(\bm{F}_\psi)) \\[10pt] 
&= \ds \frac{1}{2}K_\psi\left(\frac{1}{k_\psi}\left(e^{k_\psi\left(J_\psi-1\right)}-1\right) - \text{ln}(J_\psi) \right) + \frac{1}{2}G_\psi\left(J_\psi^{-2/3}\text{tr}({\bm{C}}_\psi) - 3 \right),
\end{array}
\label{Wshape}
\end{equation}
where $\bm{C}_\psi(\bm{F}_\psi) = \bm{F}^T_\psi\bm{F}_\psi$, $J_\psi(\bm{F}_\psi) = \text{det}(\bm{F}_\psi)$ and $K_\psi$, $G_\psi$ and $k_\psi=10$ are filter parameters. The above fictitious energy density was proposed by \citet{dahlberg2025simultaneous} and is tailored to avoid self penetration and excessive volume change and promote designs with good mesh quality and smooth design boundaries.

\section{FE-formulations}
We discretize the equilibrium equations in Eqs. \eqref{deltaW01} and \eqref{PDEpot02} using finite elements. The discretization procedure of the hyperelastic problem is described in Section \ref{sec:Hyperelastic PDE}, whereas the shape optimization problem is considered in Section \ref{sec:Shape PDE}.

\subsection{Hyperelastic PDE}\label{sec:Hyperelastic PDE}
We discretize Eq. \eqref{deltaW01} using 10-1 tetrahedral mixed $\bm{u}$-$p$ finite elements. In our total Lagrangian FE approach, we interpolate the displacement field in each finite element $e$ using the shape functions (see e.g. \citet{bathe2006finite}), i.e. $\bm{u}(\bm{X}) \approx \bs{N}(\bm{X})\bs{u}^e$, where $\bs{u}^e$ contains the nodal displacement components, whereas the independent pressure field is interpolated as element piece-wise uniform, i.e. $\tilde{p}(\bm{X}) \approx \tilde{\textsf{p}}^e$ for $\bm{X}\in\Omega^e$ where $\tilde{\textsf{p}}^e$ is the element pressure. The virtual strains are discretized as $\delta\bs{E} = \bs{B} \delta\bs{u}^e$, where the explicit format for the discrete strain operator $\bs{B}$ appear in e.g. \citet{bathe2006finite}.

Introducing the aforementioned discretization and using the arbitrariness of $\delta\bs{u}$ and $\delta\tilde{\bs{p}}$ in Eq. \eqref{deltaW01} yields the nonlinear residual equation
\begin{equation}
\bs{r}_{{a}}(\bs{a}, \lambda) =  \begin{bmatrix}
\ds\bs{r}_{{u}}  \\[10pt]
\ds \bs{r}_p 
\end{bmatrix} = \begin{bmatrix}
\fesum\Bigg(\ds\int_{\Omega^e} \bs{B}^T\widetilde{\bs{S}} \, dV +\lambda\int_{\partial\Omega^{p,e}_{c}} \hat{p} \bs{N}^T \bm{n}_c \, dS \Bigg) \\[10pt]
\ds\fesum\int_{\Omega^e}\frac{1}{K}(p-\tilde{p}) \, dV \end{bmatrix} 
= \bs{0},
\label{res}
\end{equation}
where $\fesum$ is the finite element assembly operator and we introduce $\bs{a}=[\bs{u},\ \tilde{\bs{p}}]^T$ and the load scaling parameter $\lambda\in[0,1]$. 

During loading, the equilibrium response in Eq. \eqref{res} can exhibit non-linear phenomena such as snapping and bifurcations. Therefore we use an arclength method, wherein we explicitly enforce the auxiliary hypersurface constraint
\begin{equation}
r_c = \bm{\Delta}\bs{a}^T\bm{\Delta}\bs{a} - \alpha^2 = 0,
\label{arclength}
\end{equation}
in every loadstep (\citet{crisfield1997non}). In the above, $\alpha > 0$ is a parameter that controls the magnitude of the load increment and $\bm{\Delta}\bs{a} = \bs{a}^{(n)} - \bs{a}^{(n-1)}$, where the superscript $n$ represents the load step. In every load step, we linearize Eq. \eqref{res} about the current iterate $\bs{a}$ and $\lambda$
\begin{equation}
\bs{r}_{{a}}(\bs{a} + d\bs{a}, \lambda + d\lambda)\approx \bs{r}_{{a}}(\bs{a}, \lambda) + \frac{\partial \bs{r}_{{a}}}{\partial \bs{a}}d\bs{a} + \frac{\partial \bs{r}_{{a}}}{\partial \lambda}d\lambda = \bs{0},
\label{Newt1}
\end{equation}
and solve for the displacement and load increments $d\bs{a}$ and $d\lambda$ via
\begin{equation}
\frac{\partial \bs{r}_{{a}}}{\partial \bs{a}}d\bs{a} + \frac{\partial \bs{r}_{{a}}}{\partial \lambda}d\lambda = \bs{K}  d\bs{a} + \bs{P}d\lambda = -\bs{r}_{{a}}, \quad \text{where} \quad \bs{K} = \begin{bmatrix}
\bs{K}_{{uu}} & \bs{K}_{{u}p} \\[10pt]
\bs{K}_{p{u}} & \bs{K}_{pp} 
\end{bmatrix} \quad \text{and} \quad \bs{P} = \begin{bmatrix}
\bs{P}_{u} \\[10pt]
\bs{0} 
\end{bmatrix}.
\label{Newt2}
\end{equation}
Ultimately, we solve two linear systems $\bs{K}  d\bs{a}_r = - \bs{r}_a$ and $\bs{K}  d\bs{a}_f = -\bs{P}$ and let $d\bs{a} = d\bs{a}_r + d\lambda d\bs{a}_f$. We then obtain $d\lambda$ via algebraic manipulations of Eq. \eqref{arclength}; for details we refer to \citet{crisfield1997non}.
In Eq. \eqref{Newt2} we introduce
\begin{equation}
\begin{array}{ll}
\ds\bs{K}_{{uu}} = \fesum\int_{\Omega^e} \left(\bs{B}^T\widetilde{\bs{D}}\bs{B} + \bs{G}^T\widetilde{\bs{Y}}\bs{G} \right) \, dV +
\lambda \bs{K}^p, \\[10pt]
\ds\bs{K}_{{u}p} = \ds\bs{K}_{p{u}}^{T} = -\fesum\int_{\Omega^e} \bs{B}^TJ\bs{C}^{-1}\, dV,
\\[10pt]
\ds\bs{K}_{pp} = -\fesum\int_{\Omega^e} \frac{1}{K}\, dV,\\[10pt]
\ds\bs{P}_{u} = \int_{\partial\Omega^{p,e}_{c}} \hat{p} \bs{N}^T \bm{n}_c \, dS,
\end{array}
\label{LinMat}
\end{equation}
where $\widetilde{\bs{D}}$ is the material tangent tensor $\mathbb{D} = 4 \frac{\partial^2 \widetilde{W}}{\partial \bm{C}\partial \bm{C}}$ expressed in Voigt notation and the explicit formats of $\widetilde{\bs{Y}}$ and $\bs{G}$ appear in e.g. \citet{bathe2006finite}. The pressure load tangent contribution $\bs{K}^p$ is defined in the Appendix. We emphasize that our tangent stiffness matrix is symmetric, since the pressure is uniform and is applied to a sufficiently constrained boundary surface (\citet{mok1999algorithmic} and \citet{rumpel2004hydrostatic}). 
\subsubsection{Static condensation}
Ultimately, we do not solve the full systems in Eq. \eqref{Newt2} explicitly, but instead utilize the piece-wise discontinuous interpolation of the pressure field such that $d\tilde{{\textsf{p}}}^e$ is statically condensed out on the element level. To illustrate this procedure, consider the elementwise expansion of $\bs{K}  d\bs{a}_r = - \bs{r}_{a}$, i.e.
\begin{equation}
\begin{bmatrix}
\bs{K}_{{uu}}^e & \bs{K}_{{u}p}^e \\[10pt]
\bs{K}_{p{u}}^e & {K}_{pp}^e 
\end{bmatrix}\begin{bmatrix}
d\bs{u}^e\\[10pt]
d\tilde{\textsf{p}}^e
\end{bmatrix} = -\begin{bmatrix}
\bs{r}_{{u}}^e\\[10pt]
{r}_p^e
\end{bmatrix}.
\label{statCond0}
\end{equation}
The second row of Eq. \eqref{statCond0} reveals
\begin{equation}
d\tilde{{\textsf{p}}}^e = -\left(\textsf{K}_{pp}^e\right)^{-1}\left(\textsf{r}_p^e + \bs{K}_{p{u}}^e d\bs{u}^e\right),
\label{statCond1}
\end{equation}
which, when inserted in the first row of Eq. \eqref{statCond0}, yields
\begin{equation}
\bs{K}^e d\bs{u} = \left(\bs{K}_{{uu}}^e - \bs{K}_{{u}p}^e \left(\textsf{K}_{pp}^e\right)^{-1}\bs{K}_{p{u}}^e\right)d\bs{u}^e = -\bs{r}_{{u}}^e  + \bs{K}_{{u}p}^e\left(\textsf{K}_{pp}^e\right)^{-1}{r}_p^e=\bs{r}^e.
\label{statCond2}
\end{equation}
where $\bs{K}^e$ and $\bs{r}^e$ are the condensed element stiffness matrix and residual vector, respectively. We solve $\bs{K}  d\bs{a}_p = -\bs{P}$ using the same methodology. 

\subsection{Shape filter PDE}\label{sec:Shape PDE}
We use 10 node tetrahedral single field finite elements to solve Eq. \eqref{PDEpot02}. Again, the vector fields are interpolated via the shape functions $\bs{N}$, such that $\bm{d}(\bm{X}_o) \approx \bs{N}(\bm{X}_o)\bs{d}^e$ and $\bm{\psi} (\bm{X}_o)\approx \bs{N}(\bm{X}_o)\bm{\uppsi}^e$, where $\bs{d}^e$ and $\bm{\uppsi}^e$ are the element nodal shape displacement and filtered shape displacement vectors. Using the arbitrariness of $\delta\bm{\psi}$, the discretized version of Eq. \eqref{PDEpot02} requires $\bm{\uppsi}$ to satisfy
\begin{equation}
 \bs{r}_\psi(\bm{\uppsi};\bs{d}) = \fesum\left(\ds\int_{\Omega_o^e} \bs{B}^T\bs{S}_\psi \, dV - \ds\int_{\partial\Omega_o^{d,e}} \bs{N}^T\left(\bm{d}-\bm{\psi}\right) \, dS \right)= \bs{0},
\label{PDE04}
\end{equation}
To solve Eq. \eqref{PDE04}, we utilize a traditional Newton's method. For more details we refer to \citet{dahlberg2025simultaneous}.

\section{Optimization problem}
We use our shape optimization framework to design soft actuators with bespoke deformation patterns. To this end, our multiobjective shape optimization problem reads
\begin{equation}
(\mathbb{SO}) \ \begin{cases}
\underset{\bs{d}}{\text{min}} \ \underset{i \in[1,n_g]}{\text{max}} \ \ \{g_i\}, \\[10pt]
\text{s.t} \quad \begin{cases}
f \leq 0, \quad \\
\textsf{d}_k\in[\underline{\textsf{d}}, \overline{\textsf{d}}], \quad k\in[1, 3n_n],
\end{cases}
\end{cases}
\label{opt}
\end{equation}
where $n_g$ is the number of objectives, $n_n$ is the number of finite element nodes and the design shape variables are constrained to be in the range of $\underline{\textsf{d}} \leq \textsf{d}_j \leq \overline{\textsf{d}}$. The range of the box constraints is a delicate choice made by the user. In the above, $g_i$ are objective functions and $f$ is an inequality constraint. In our numerical experiments, we follow \citet{dahlberg2025simultaneous} and use rather tight bounds, but increase the design freedom by performing several spaced updates of the mesh coordinates such that $\bm{X}_o \leftarrow \bm{X}_o + \bm{\psi}$ and $\bs{d} \leftarrow \bm{0}$. Our experience is that this approach produces designs with better mesh quality compared to those designs obtained using large bounds without mesh coordinate updates. We emphasize that the equilibrium equality constraints $\bs{r}_{{a}} = \bs{0}$ and $\bs{r}_{\psi} = \bs{0}$ are explicitly enforced, i.e. we solve the optimization problem in a staggered fashion. The gradient-based nonlinear programming method MMA is used to solve the optimization problem defined in Eq. \eqref{opt}, and we utilize its min-max formulation (\citet{svanberg1987method}). 

We seek designs of soft actuators which exhibits desired deformation behaviors over a surface $\partial\Omega^d\subset \partial\Omega$ at target pressures. To this end, each objective is defined as
\begin{equation}
g_i = \frac{w_i}{\vert\partial\Omega^d\vert}\int_{\partial\Omega^d} \bm{u}^{i} \cdot \bm{v}^i \, dS,
\label{g0}
\end{equation}
where $w_i$ are weights, $\vert\partial\Omega^d\vert = \int_{\partial\Omega^d} \, dS$ is the area, $\bm{u}^{i}$ is the displacement field and $\bm{v}^i$ is the vector direction field at target point $i$. In Eq. \eqref{opt}, we also introduce the inequality constraints 
\begin{equation}
f = \frac{1}{\vert\partial\Omega^d\vert}\int_{\partial\Omega^d} \vert\bm{u}^{j} \cdot \bm{s}^j \vert^2 \, dS - \epsilon_u \leq 0,
\label{constraint}
\end{equation}
which constrains the deformation at target point $j$ in the direction $\bm{s}^j$ to be smaller than the relaxation tolerance $\epsilon_u \geq 0$. By solving the optimization problem as posed in Eq. \eqref{opt}, we promote designs which exhibit tailored deformation patterns.

\subsection{Sensitivity analysis}
We compute the gradients of a function $\tilde{g}(\bs{d}) =g\left({\bm{\uppsi}}(\bs{d}),\bs{a}({\bm{\uppsi}}(\bs{d}))\right)$ using the adjoint method. In this method, we augment $\tilde{g}$ with the equality constraints ($\bs{r}_{{a}} = \bs{0}$ and $\bs{r}_{{{\psi}}} = \bs{0}$) via the associated Lagrange multipliers $\bm{\mu}_{{a}}$ and $\bm{\mu}_{{{\psi}}}$ to obtain the identity
\begin{equation}
\bar{g} := \tilde{g} - \bm{\mu}_{{a}}^T\bs{r}_{{a}} - \bm{\mu}_{{{\psi}}}^T\bs{r}_{{{\psi}}}.
\label{sens01}
\end{equation}
Next, we differentiate Eq. \eqref{sens01} with respect to $\bs{d}$ and rearrange, to obtain
\begin{equation}
\ds\frac{d\bar{g}}{d \bs{d}} = - \bm{\mu}_{{{\psi}}}^{T}\frac{\partial \bs{r}_{{{\psi}}}}{\partial \bs{d}}+ \Bigg[\frac{\partial g}{\partial \bm{\uppsi}} - \bm{\mu}_{{a}}^T\frac{\partial \bs{r}_{{a}}}{\partial {\bm{\uppsi}}}  - 
\bm{\mu}_{{{\psi}}}^{T}\frac{\partial \bs{r}_{{{\psi}}}}{\partial {\bm{\uppsi}}}  +\left(\frac{\partial g}{\partial \bs{a}}- \bm{\mu}_{{a}}^T\frac{\partial \bs{r}_{{a}}}{\partial \bs{a}}  \right)\frac{\partial \bs{a}}{\partial {{\bm{\uppsi}}}} \Bigg]\frac{\partial {\bm{\uppsi}}}{\partial \bs{d}}.
\label{sens03}
\end{equation}
To annihilate the implicit derivatives $\frac{\partial \bs{a}}{\partial {{\bm{\uppsi}}}}$ in Eq. \eqref{sens03}, we solve the linear system
\begin{equation}
\left(\frac{\partial \bs{r}_{{a}}}{\partial \bs{a}}\right)^T \bm{\mu}_{{a}} = \left(\frac{\partial g}{\partial \bs{a}}\right)^T,
\label{adj2}
\end{equation}
to obtain $\bm{\mu}_{{a}}$. We emphasize that we again use static condensation when solving the above linear system. Inserting Eq. \eqref{adj2} in Eq. \eqref{sens03} gives
\begin{equation}
\ds\frac{d\bar{g}}{d \bs{d}} = \ds - \bm{\mu}_{{{\psi}}}^{T}\frac{\partial \bs{r}_{{{\psi}}}}{\partial \bs{d}} 
 + \Bigg[\frac{\partial g}{\partial {{\bm{\uppsi}}}}  - \bm{\mu}_{{a}}^T\frac{\partial \bs{r}_{{a}}}{\partial {{\bm{\uppsi}}}} 
- \bm{\mu}_{{\psi}}^{T}\frac{\partial \bs{r}_{{{\psi}}}}{\partial {{\bm{\uppsi}}}}
\Bigg]\frac{\partial {{\bm{\uppsi}}}}{\partial \bs{d}}.
\label{sens04}
\end{equation}
The implicit derivative $\frac{\partial {{\bm{\uppsi}}}}{\partial \bs{d}}$ is annihilated by solving the adjoint problem
\begin{equation}
\left(\frac{\partial \bs{r}_{{{\psi}}}}{\partial {\bm{\uppsi}}}\right)^T  \bm{\mu}_{{{\psi}}} = \left(\frac{\partial g}{\partial {\bm{\uppsi}}}\right)^T  - \left(\frac{\partial \bs{r}_{{a}}}{\partial {\bm{\uppsi}}}\right)^T\bm{\mu}_{{a}},
\label{adj3}
\end{equation}
for $\bm{\mu}_{{{\psi}}}$. Finally, the sensitivity expression reduces to
\begin{equation}
\ds\frac{d \bar{g}}{d \bs{d}} = - \bm{\mu}_{{{\psi}}}^{T}\frac{\partial \bs{r}_{{{\psi}}}}{\partial \bs{d}}.
\label{sens06}
\end{equation}
We emphasize that since we utilize an arclength method to solve the equilibrium equations, care must be taken to ensure that we always compute the sensitivity at set target pressures. To this end, we monitor the pressure during loading, and when we surpass a target pressure, we return to the previous load step and perform a standard force-controlled Newton step to exactly end up at the set target pressure. More details for the sensitivity computations are provided in the Appendix.

\section{Experiments}
To validate the proposed shape optimization framework, we fabricate and test several of the inversely designed soft actuators. The fabrication and experimental testing procedure are described in detail in the following two sections. 

\subsection{Fabrication}
We cast our soft actuators using polyvinyl siloxane (PVS, Zhermack Elite Double 32). For each tested design, we fabricate and test two actuators to ensure repeatability. The molds are design in Blender based on the surface representations (STL) of our optimized designs, which we extract using ParaView. We adopt a compound molding technique in which the molds consists of rigid positive and flexible negative parts. The rigid positive molds are 3D printed using the Formlabs Form 3B with 0.05 mm layer thickness using the Model V3 resin. After the print is finished, the molds are washed in isopropanol in the Formlabs Wash V1 for 10 minutes to remove excess resin. Subsequently, any support structure is manually removed, after which the part is washed for another 5 minutes. The molds are then cured at room-temperature for at least 24 hours.

\begin{figure}[H]
    \centering
    \includegraphics[width=\textwidth]{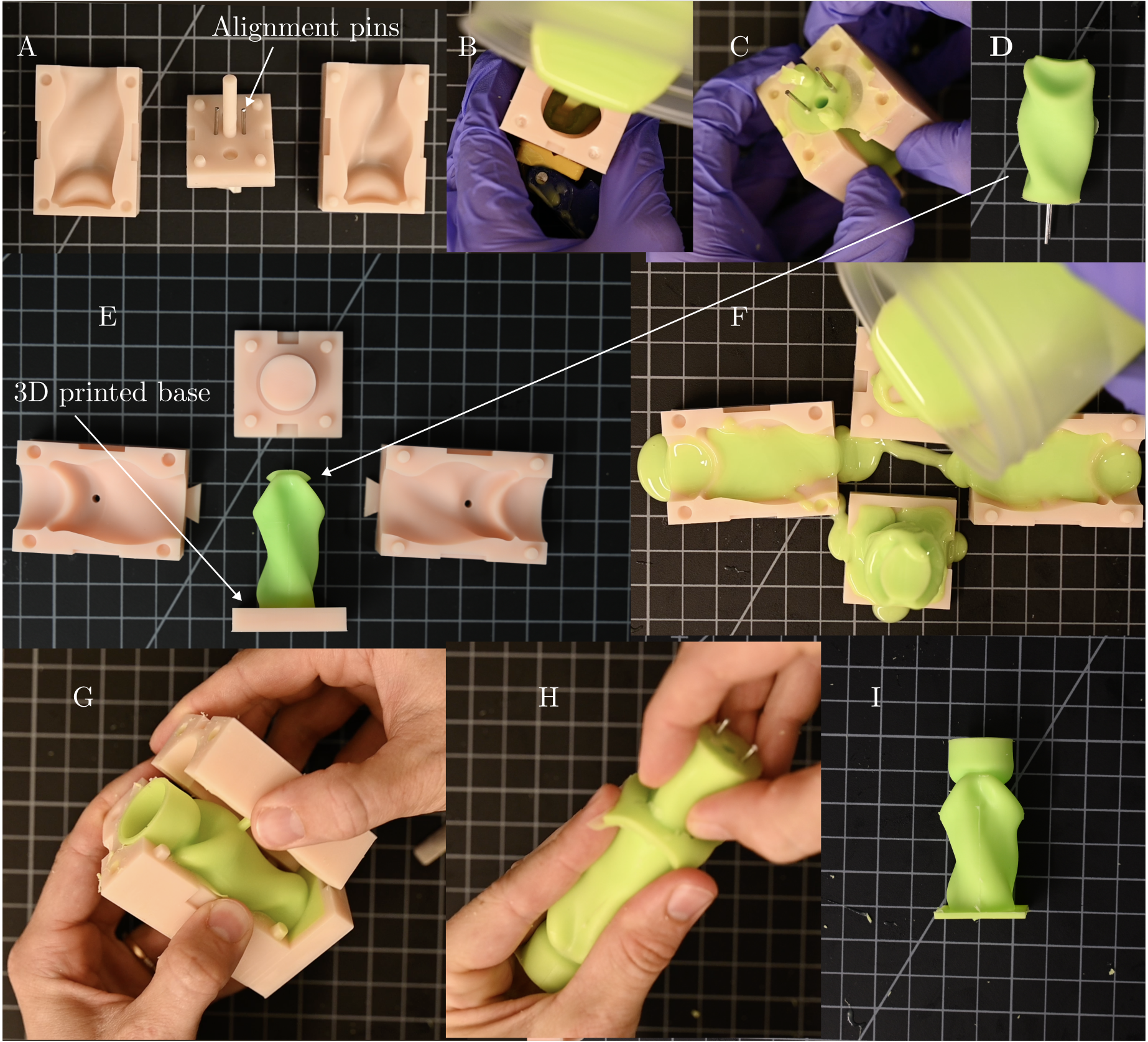}
    \caption{\textbf{Fabrication}. The two-step molding process. }
    \label{fig:setup}
\end{figure}

The soft actuator are casted via a two-step molding process, as illustrated in Fig. \ref{fig:setup}. First, inspired by the work of \citet{galloway2016soft}, we cast the flexible negative mold from the PVS. 3D printed molds are used to cast the complementary soft negative mold of the actuator's closed cavity geometry using PVS. Before casting, we insert two metal rods in the bottom of the mold which will act as alignment pins when mounting, and we apply a layer of mold release agent (Mann Ease Release 200) on the positive mold  (Fig. \ref{fig:setup}A). Then we mix the two-part, base and catalyst, PVS elastomer (THINKY mixer ARE-310) at 2000 RPM for 30 seconds, and subsequently at 2200 RPM for 30 seconds, to ensure uniformity in its composition and to degas. We then pour the mixed PVS inside the mold (Fig. \ref{fig:setup}B), which is clamped and placed at room temperature for 20 minutes before opening it. After demolding (Fig. \ref{fig:setup}C), the negative flexible mold (with alignment pins, see Fig. \ref{fig:setup}D) is mounted on the rigid positive base plate of the actual actuator mold (Fig. \ref{fig:setup}E), thereby defining the external geometry of the actuator to form the final casting assembly (Fig. \ref{fig:setup}E). We do not bake the flexible negative mold.  We acknowledge that the compliance of the PVS mold introduces small manufacturing errors, but have found that for intricate cavity shapes, the de-molding might otherwise cause severe rupture of the actuator, wherefore a soft negative mold proves beneficial.


The second step is to cast the actual actuators. First, we apply a layer of mold release agent (Mann Ease Release 200) to both the positive and negative molds. Then we mix PVS elastomer using the same technique as explained above. The mixed PVS is then poured inside the mold (Fig. \ref{fig:setup}F), which is clamped and placed at room temperature for 20 minutes. The demolding process consist of first removing the positive mold (Fig. \ref{fig:setup}G), and then carefully peeling the actuator from the negative mold (Fig. \ref{fig:setup}H). The actuators (Fig. \ref{fig:setup}I) are subsequently baked at $60$ $\degree$C for 24 hours in an oven to stabilize the mechanical properties.

The actuators are finally mounted on a laser-cut (Universal Laser Systems PLS6.150D with a 75W C02 lase) acrylic support plate, which is engraved to increase roughness and promote bonding. To do this, we first apply a primer (Aron Alpha PP Primer) to the base of the actuator and to the acrylic support plate to promote adhesion. Then we use Locktite SuperGlue to glue the actuator to the acrylic plate. The glue is left to cure for 24 hours.

\subsection{Testing}
\begin{figure}[H]
    \centering
    \includegraphics[width=\textwidth]{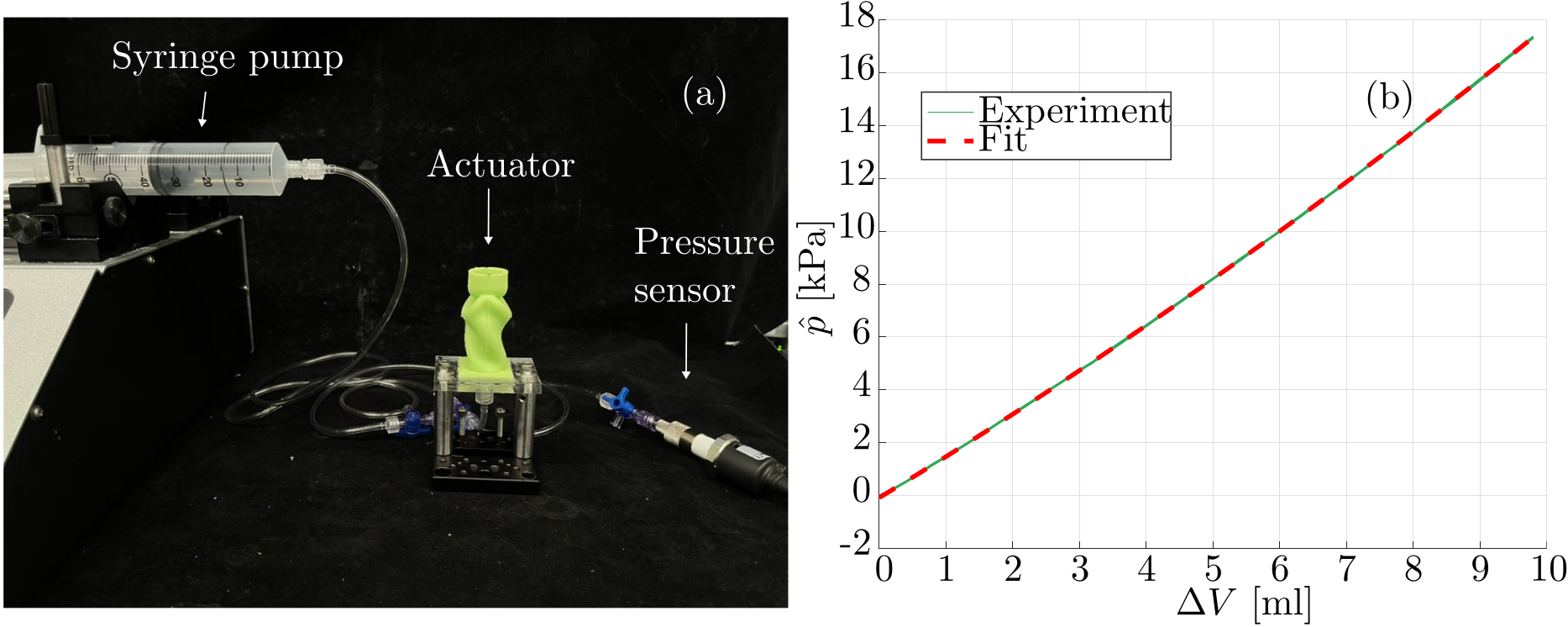}
    \caption{\textbf{Testing}. \textbf{(a)} The testing setup. \textbf{(b)} The pressure-volume relation of the system without the actuator. }
    \label{fig:compliance}
\end{figure}

The actuators are tested using a   syringe pump (Harvard Apparatus 33DS), pressure sensor (15 PSI Ashcroft GV Pressure Transducer), data acquisition device (DAQ, Saleae Logic pro 8), tubing, and mount for the actuator (Fig. \ref{fig:compliance}(a)). Using the syringe pump, we perform volume-controlled loading of air with a flow rate of $10$ ml$/$min. Both the pressure signal and volume input are recorded as a function of time using the DAQ.  During the tests, actuator deformation was recorded using a  digital camera (Nikon Z 6II). Black dot markers were placed on the actuator surface, and their positions were tracked in the recorded videos using the Kanade-Lucas-Tomasi (KLT) feature-tracking algorithm implemented in MATLAB's Computer Vision Toolbox. 

To compare the numerically predicted and experimentally measured pressure–volume curves, it is necessary to account for both the compliance of the experimental setup and the compressibility of air. The compliance of the setup, which originates from the tubing and syringe, is quantified by measuring the pressure–volume response of the system without the actuator. Specifically, 10 mL of air is infused at a constant flow rate of 10 mL/min, and the corresponding pressure response is recorded. The measured pressure–volume curve, along with a quadratic fit, is presented in Fig.~\ref{fig:compliance}(b). The quadratic fit can be extrapolated to provide the volume correction due to the system compliance, $\Delta V_{compliance}$, such that 
\begin{equation}
    \Delta V_{corr} = \Delta V_{syringe} - \Delta V_{compliance},
    \label{compliance}
\end{equation}
where $\Delta V_{syringe}$ is the amount of volume dispensed by the syringe and $\Delta V_{corr}$ is the system compliance corrected volume change.

The compressibility of air is accounted for under two assumptions: (i) the flow rate is sufficiently low for the process to be considered isothermal, and (ii) the enclosed air obeys the ideal gas law. Under these assumptions, the volume change of the pressurized cavity of the actuator, $\Delta V^p$, can be expressed as (see, e.g., \citet{yang2024complex,gorissen2020inflatable} for details)
\begin{equation}
\Delta V^p = \Delta V_{corr} - \frac{\hat{p}}{\hat{p}+\hat{p}_0} V^p,
\label{dV}
\end{equation}
where $\hat{p}_0$ denotes the atmospheric pressure and $V^p$ is the initial cavity volume. Throughout this study, the volume change of the pressurized cavity in our experiments is computed using Eq.~\eqref{dV}. 

\section{Results}
We demonstrate our shape optimization framework by morphing the initial geometry depicted in Fig.~\ref{fig:cylinder} to achieve a range of target deformations. The initial geometry consists of a closed lower (blue) cylindrical cavity with an open (gray) cylindrical cavity positioned on top. The cylinder is clamped at the lower surface $\partial\Omega^u$. Pressurized gas is injected into the lower closed cylindrical cavity defined by  the surface $\partial\Omega^p$. We target various deformation modes of the open surface $\partial\Omega^d$ of the top open cylinder by employing the objectives and constraints defined in Eqs.~\eqref{g0} and \eqref{constraint}. The target deformation modes include extension, contraction, and grasping through closure of the top cylindrical cavity. The geometrical parameters defining the initial geometry are summarized in Table~I.


To solve general 3D design problem, we require parallel computing in the form of message passing (MPI) routines and linear algebra tools from PETSc (\citet{balay2019petsc}). We generate the unstructured finite element grids using Gmsh (\citet{geuzaine2009gmsh}), and the discretizations are distributed using ParMETIS (\citet{karypis1998parallel}) and are managed by DMPlex structures in PETSc. The linear systems arising from the discretized PDE:s are solved using the direct solver MUMPS (\citet{amestoy2000mumps}), which provides the Cholesky factorizations. All numerical solutions are obtained using the COSMOS cluster at Lund University and are run on one node consisting of two AMD EPYC 7413 processors (48 CPU cores @ 3.6 GHz) and 256 GB of ram.

In all numerical examples, we set the MMA move limit to $0.025$. The optimization is terminated after 350 design iterations, since we after this observe minimum changes in the design and objective. The optimizer is restricted from morphing the shape of the actuator at the attachment surface, $\partial \Omega^u$, and at the top cylindrical cavity surface, $\partial \Omega^d$, i.e. $\partial\Omega_o^\psi = \partial \Omega^u \cup \partial \Omega^d$. The finite element mesh consists of $109604$ quadratic tetrahedral elements, resulting in $n_n = 187611$ nodes. The material properties of the PVS are $E = 1.2$ MPa and $\nu = 0.49$ (\citet{yang2024complex}). We omit the constraint $f$ in the optimization problem definition in Eq. \eqref{opt}, unless stated otherwise. 

We emphasize that the choices of $K_\psi$, $G_\psi$, $\overline{\textsf{d}}$ and $\underline{\textsf{d}}$ are problem dependent and may require some numerical investigation. For example, large bounds on the box constraints on the design variables $\overline{\textsf{d}}$ and $\underline{\textsf{d}}$ and small $K_\psi$ and $G_\psi$ modulii enables large shape modifications, with the risk of producing designs with tangled meshes and overlapping surfaces. Decreasing the bounds and increasing the modulii mitigates this effect, at the expense of limiting the extent of the shape modifications. Indeed, a trade-off between allowing larger shape changes and ensuring a reasonable mesh quality exists. In the following examples, we fix $K_\psi = 1 \times 10^{-5}$ mm$^{-1}$ and $\overline{\textsf{d}} = -\underline{\textsf{d}} = 1.2$ mm, but alter $G_\psi$ in Eq. \eqref{Wshape} to accommodate the problem specific mesh quality.

\begin{center}
\begin{minipage}{0.45\textwidth}
    \centering
    \includegraphics[width=0.7\textwidth]{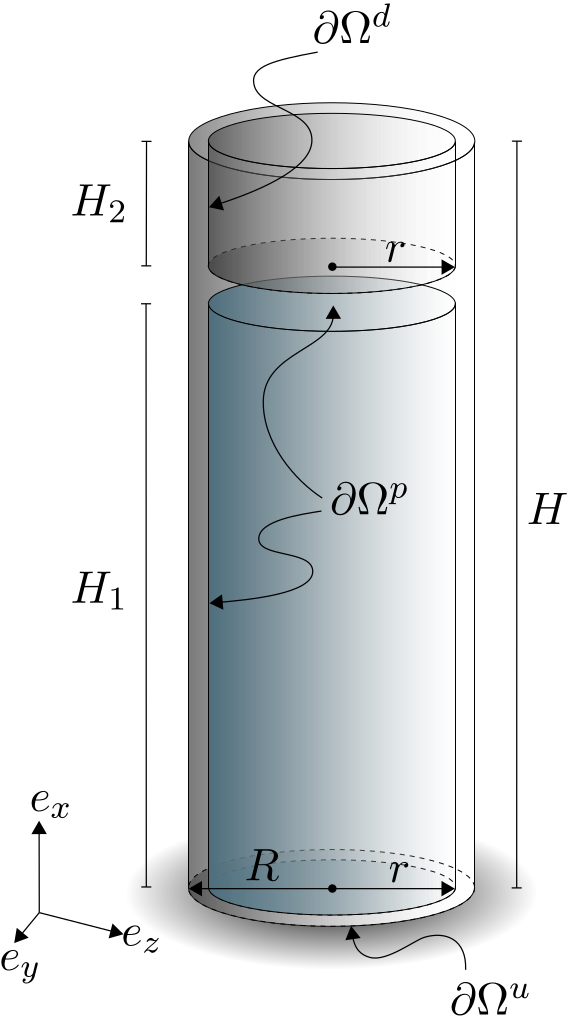}
    \captionof{figure}{\textbf{Geometry}. The initial geometry of the actuator.}
    \label{fig:cylinder}
\end{minipage}
\hfill
\begin{minipage}{0.5\textwidth}
    \centering
     \begin{tabular}{lr}
        \hline
        Inner radius $r$ &  $10$\\
        Outer radius $R$ &  $11.5$  \\ 
        Total height $H$ &  $60$ \\ 
        Pressurized cavity height $H_1$ &  $47$\\
        Top cavity height $H_2$ &  $10$\\
        \hline
        \end{tabular}
        \label{tab:geom}
    \captionof{table}{The geometrical parameters defining the initial geometry, all in millimeters.}
\end{minipage}

\end{center}

\subsection{Inverse design of a gripping actuator}
\begin{figure}[H]
    \centering
    \includegraphics[width=0.25\textwidth]{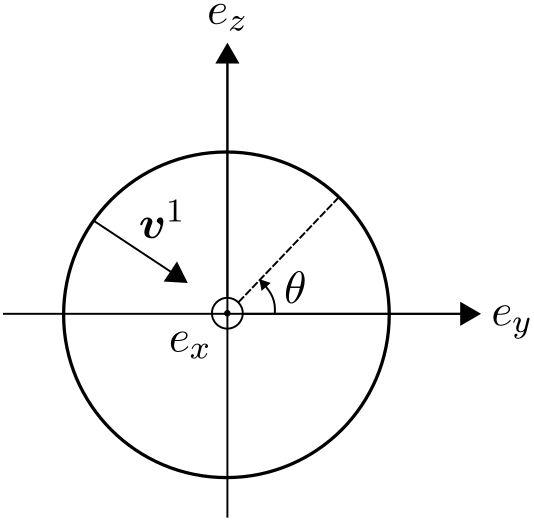}
    \captionof{figure}{\textbf{Target direction}. The definition of the negative radial direction $\bm{v}^1$ for the gripping actuation mechanism.}
    \label{fig:circle}
\end{figure}

In the first example, we design an actuator capable of grasping an object by maximizing the inward radial displacement of the cylindrical surface $\partial\Omega^d$ shown in Fig.~\ref{fig:cylinder} upon application of a pressure $\hat{p} = 15$~kPa to the closed cavity. To this end, we minimize the objective function $g_1$ defined in Eq.~\eqref{g0} with $\bm{v}^1 = (0, -\cos(\theta), -\sin(\theta))$ and $w_1 = -1$, where $\theta$ denotes the angular coordinate in the $e_y$-$e_z$ plane (Fig.~\ref{fig:circle}). For this problem, we consider a single objective, which is computed at the maximum pressure, i.e. at a load scaling factor of $\lambda = 1$ (see Eq. \eqref{res}). We update the mesh coordinates every 50 design iterations, for a total of three updates, and use a filter parameter of $G_\psi = 1\times10^{-2}$~mm$^{-1}$ in Eq. \eqref{Wshape}.

\begin{figure}
    \centering
    \includegraphics[width=0.9\textwidth]{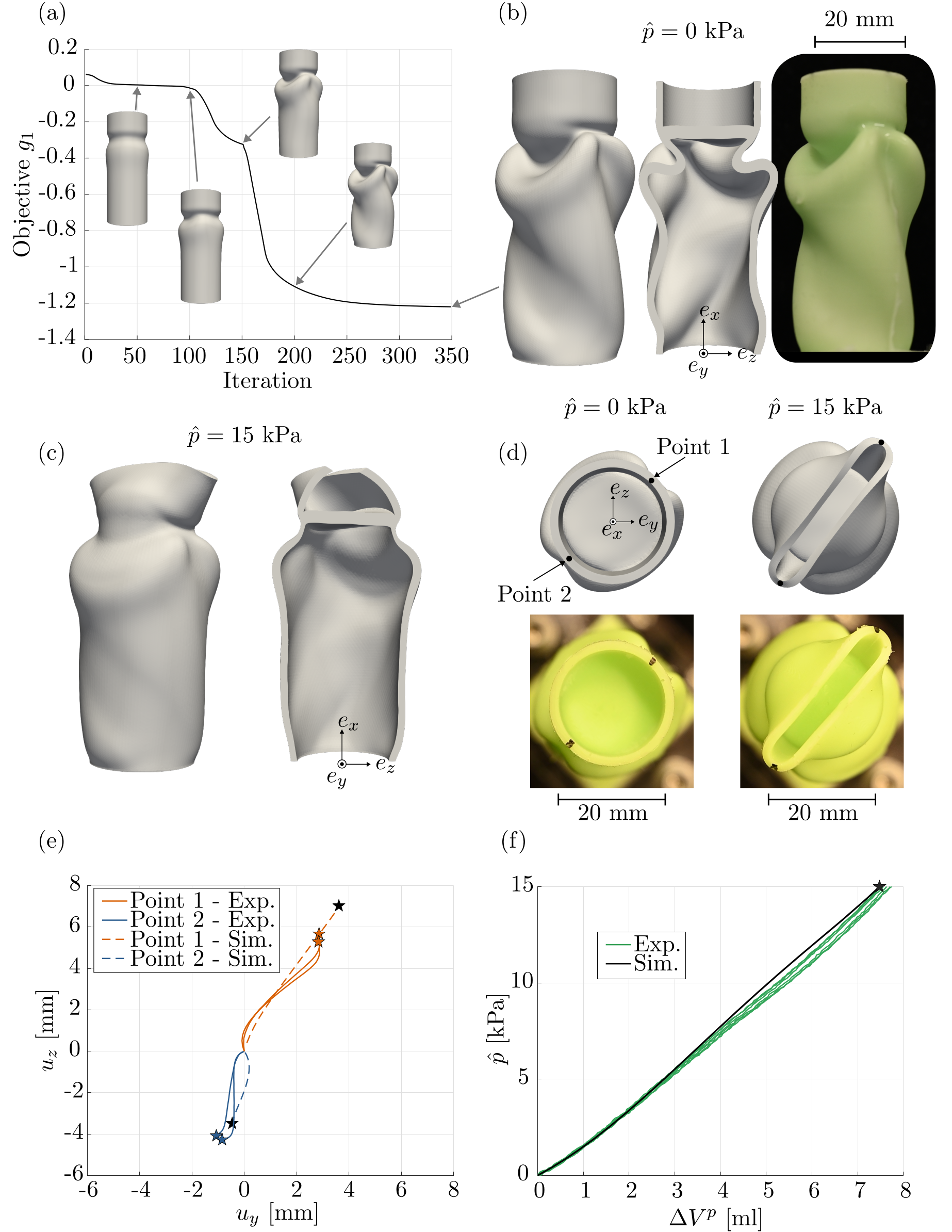}
    \captionof{figure}{\textbf{The gripping actuator}. \textbf{(a)} The evolution of the objective function $g_1$ over the design iterations, along with snapshots of intermediate designs. \textbf{(b)} The side-view and sliced-view of the design and its realization in the undeformed configuration, i.e. at $\hat{p}= 0$ kPa. \textbf{(c)} The side-view and sliced-view of the design in the deformed configuration at maximum pressure $\hat{p}= 15$ kPa. \textbf{(d)} The top-view of the design and its realization in the undeformed and deformed configurations. \textbf{(e)} The displacement of the two points illustrated in (d) during pressurization. \textbf{(f)} The applied pressure $\hat{p}$ versus cavity volume change $\Delta V^p$.}
    \label{fig:grabber}
\end{figure}

The history of the objective function evolution is plotted in Fig. \ref{fig:extender}(a), where also snapshots of the intermediate designs are included. We observe that the objective function  monotonically decreases, with drastic drops appearing in conjunction with every mesh coordinate update, as expected. The optimized design and its cast realization is depicted in Fig.~\ref{fig:grabber}(b) in its undeformed configuration at $\hat{p}=0$~kPa. The design is depicted in its deformed configuration at $\hat{p}=15$~kPa in Fig.~\ref{fig:grabber}(c). These snapshots indicate that, in the optimal design, two arch-shaped bulges beneath the top cylinder, $\partial\Omega^d$, expand outward during pressurization and induce the observed elliptical deformation pattern of $\partial\Omega^d$. 
The top views of the open cavity shown in Fig.~\ref{fig:grabber}(d) clearly indicate that the initially circular opening transforms into an elongated elliptical shape, thereby facilitating \say{gripping}. Fig.~\ref{fig:grabber}(b)-(c) also includes sliced views of the actuator in the $e_z$-$e_x$ plane, further illustrating the deformation mechanism. 

To validate the computational framework, we experimentally characterize the response of the identified optimized geometry. We perform motion tracking of the two black markers highlighted in Fig. \ref{fig:grabber}(d) and compare the experimental and simulated displacements over time (Fig. \ref{fig:grabber}(e)), see also the movie provided as the supplementary material. Although the overall behavior agrees well (see the deformation patterns in the deformed configuration in Fig. \ref{fig:grabber}(d)), we note slight discrepancies between the numerical and experimental motion tracking results. We believe that these errors stem from manufacturing errors, such as mold misalignment and air bubbles, that invariably appear in the manufacturing process. Also, camera misalignment can play a role in the discrepancies, especially since the magnitude of the displacements is relatively small. To further characterize the gripping actuator, we compare the numerically computed and experimentally measured pressure-volume relation. The result is depicted in Fig. \ref{fig:grabber}(f), where we plot the applied pressure $\hat{p}$ versus volume change of the pressurized cavity\footnote{We compute the current cavity volume $V^p_c$ using the divergence theorem, i.e. $V^p_c = \int_{\Omega^p} dV = \frac{1}{3}\int_{\Omega^p} \bm{\nabla}_c\cdot \bm{x} \, dV = \frac{1}{3}\int_{\partial\Omega^p} \bm{x} \cdot \bm{n}_c \, dS$.}, i.e. $\Delta V^p = V^p_c - V^p$. We observe a nearly linear pressure-volume relation, and a good correlation between the numerical and experimental results (the experiments were conducted three times each for the two fabricated actuators). Overall, the numerical and experimental results show good agreement, indicating that the proposed numerical framework accurately captures the behavior of the designed actuators.

\begin{figure}
    \centering
    \includegraphics[width=\textwidth]{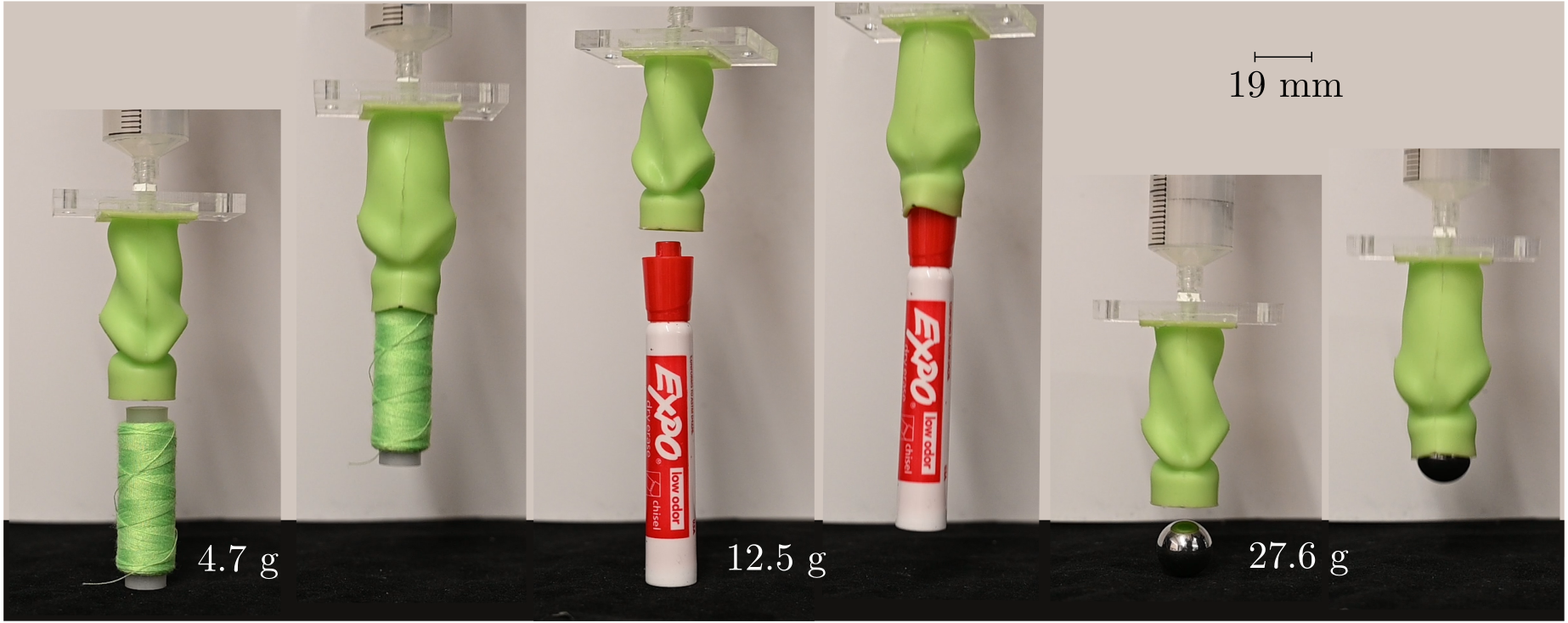}
    \captionof{figure}{\textbf{Grasping examples}. Photos of the gripping actuator grabbing and lifting three objects: a thread spool (4.7 g), a whiteboard pencil (12.5 g) and a metallic ball (27.6 g). For reference, the fabricated actuator weights 11.9 g. The actuator is manually pressurized using a syringe and lifted by hand.}
    \label{fig:grabbing}
\end{figure}

Finally, Fig.~\ref{fig:grabbing} demonstrates the capabilities of the gripping actuator through a series of grasping experiments. The actuator successfully grasps and lifts several objects, including a thread spool (4.7 g), a whiteboard marker (12.5 g), and a metallic ball (27.6 g). All objects have diameters smaller than 20 mm, requiring the actuator to actively grasp rather than simply support them. We note that the actuator is pressurized manually using a syringe and that the lifting motion is performed by hand. Despite these simple experimental conditions, the actuator is capable of reliably grasping and lifting objects of varying shapes and weights.

\subsection{Inverse design of linear actuators}
In the next example, we design actuators that maximizes the output displacement parallel to the $e_x$-direction upon pressurization to $\hat{p}=15$ kPa, thereby acting as linear actuators. We perform two optimizations to design 1) an extending actuator by setting $\bm{v}^1=(1,0,0)$ in Eq.~\eqref{g0} and 2) a contracting actuator by setting $\bm{v}^1=(-1,0,0)$. For both cases, we employ a single objective function and $w_1 = -1$, evaluated at a load scaling factor of $\lambda=1$. Furthermore, the mesh coordinates are updated every 50 design iterations, resulting in a total of three mesh updates. The filter parameter is set to $G_\psi=3\times10^{-2}$ mm$^{-1}$ in Eq.~\eqref{Wshape}. 

\begin{figure}

    \centering
    \includegraphics[width=0.8\textwidth]{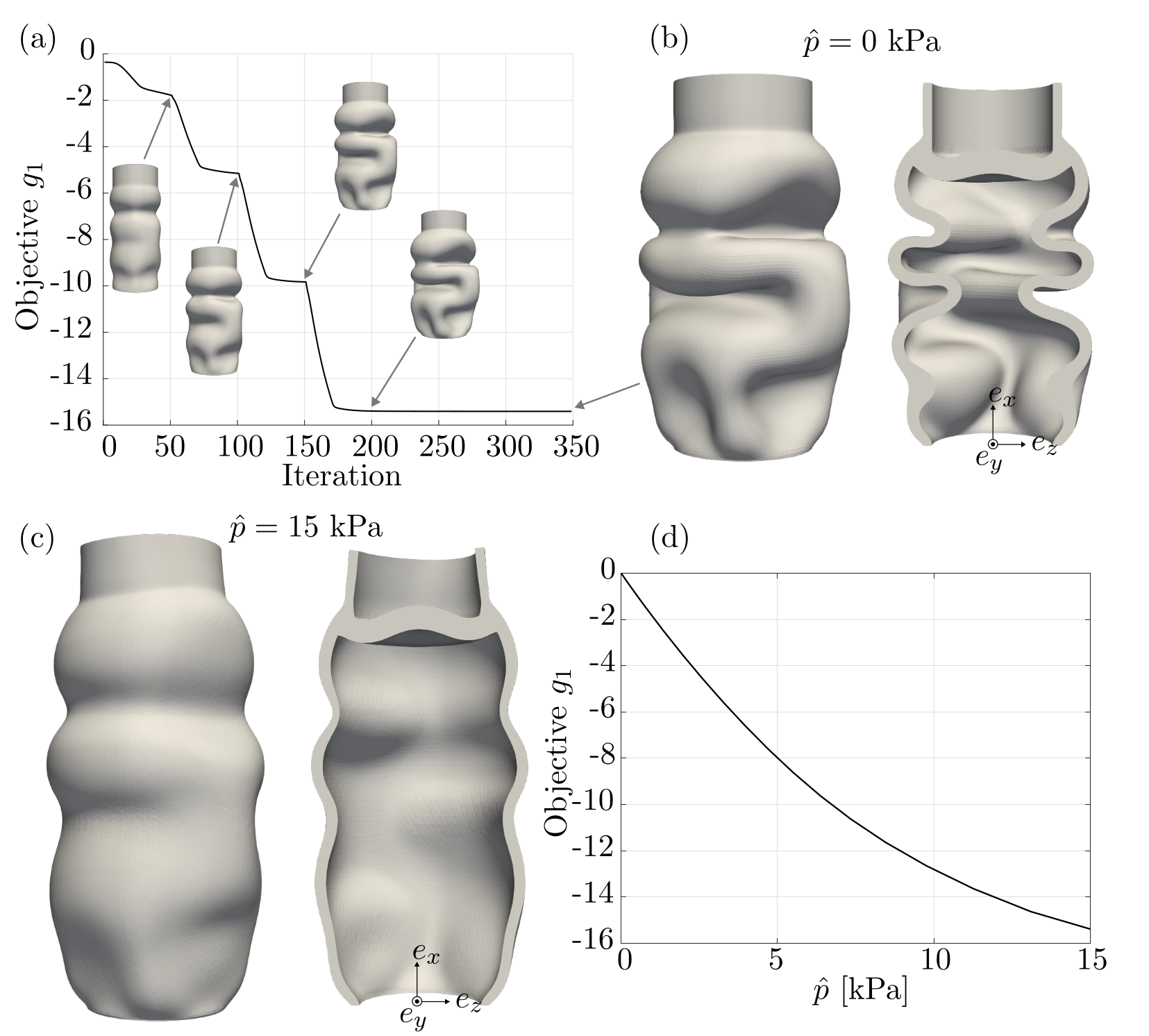}
    \captionof{figure}{\textbf{The extending linear actuator}. \textbf{(a)} The evolution of the objective function $g_1$ over the design iterations, along with snapshots of intermediate designs. \textbf{(b)} The side-view and sliced-view of the design in the undeformed configuration, i.e. at $\hat{p} = 0$ kPa. \textbf{(c)} The side-view and sliced-view of the design in the deformed configuration at the maximum pressure $\hat{p} = 15$ kPa. \textbf{(d)} The objective function $g_1$ versus the applied pressure $\hat{p}$.}
    \label{fig:extender}
\end{figure}

Results for the extending actuator are shown in Fig.~ \ref{fig:extender}. Again,  we observe a monotonically decreasing objective function $g_1$ (Fig. \ref{fig:extender}(a)). The identified optimal design and its sliced view in the $e_z$-$e_x$ plane are depicted in Fig. \ref{fig:extender}(b) in the undeformed configuration at $\hat{p} = 0$ kPa and in Fig. \ref{fig:extender}(c) in the deformed configuration at $\hat{p} = 15$ kPa. Unsurprisingly, the optimizer morphs the shape of the actuator to form bulges, which results in expansion in the positive $e_x$-direction upon pressurization. In Fig. \ref{fig:extender}(d), we plot the objective function $g_1$, which corresponds to the area average surface integral of the displacement over the surface $\partial\Omega^d$, versus the applied pressure $\hat{p}$. We find that $g_1$ monotonically decreases with increasing pressure and that the largest expansion magnitude occurs at the maximum pressure. We emphasize that the objective $g_1$ is negative, as we have a minimization optimization problem. 


\begin{figure}
    \centering
    \includegraphics[width=0.8\textwidth]{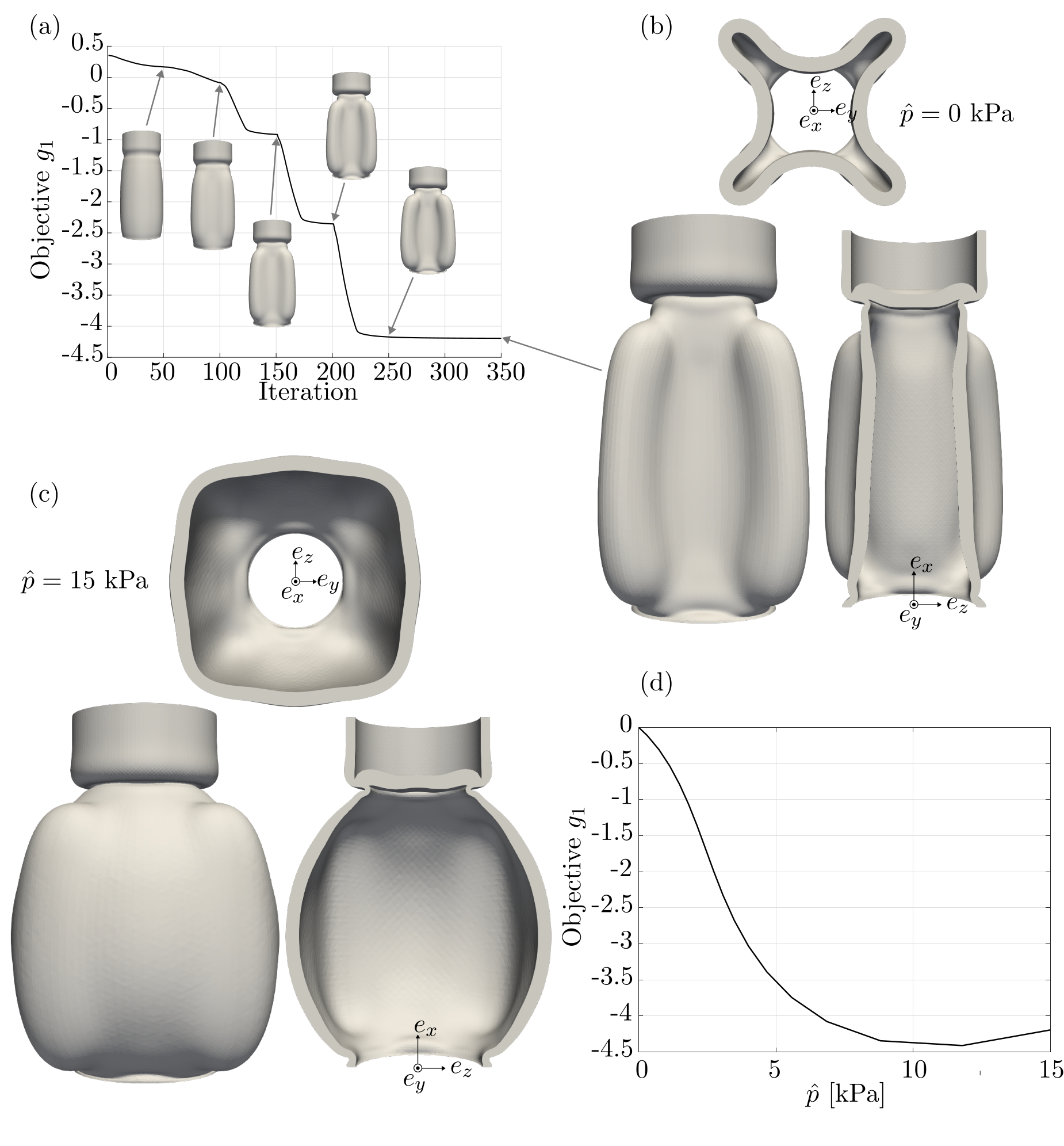}
    \captionof{figure}{\textbf{The contracting linear actuator}. \textbf{(a)} The evolution of the objective function $g_1$ over the design iterations, along with snapshots of intermediate designs. \textbf{(b)} The side-view and sliced-view of the design in the undeformed configuration, i.e. at $\hat{p} = 0$ kPa. \textbf{(c)} The side-view and sliced-view of the design in the deformed configuration at the maximum pressure $\hat{p} = 15$ kPa. \textbf{(d)} The objective function $g_1$ versus the applied pressure $\hat{p}$.}
    \label{fig:contractor}
\end{figure}

Results for the contracting actuator are shown in Fig.~\ref{fig:contractor}. The evolution of the objective function is presented in Fig.~\ref{fig:contractor}(a), together with snapshots of the intermediate designs. The final optimized design is shown in Fig.~\ref{fig:contractor}(b). The optimization morphs the geometry such that a cross-section in the $e_y$–$e_z$ plane exhibits four curved walls arranged in a cross-shaped pattern. Upon pressurization, these walls expand outward within the plane, causing the upper cylindrical section to contract in the negative $e_x$ direction, as illustrated in Fig.~\ref{fig:contractor}(c).

Interestingly, the maximum magnitude of the desired displacement occurs at a pressure below the target value of $\hat{p}=15$ kPa. This behavior is evident in Fig.~\ref{fig:contractor}(d), which shows the objective function $g_1$  as a function of the applied pressure $\hat{p}$. We attribute this phenomenon to a ballooning effect: above a certain pressure level, the initially curved walls shown in Fig.~\ref{fig:contractor}(b) progressively approach the nearly square deformation pattern depicted in Fig.~\ref{fig:contractor}(c). As a result, further pressurization leads to displacement in the opposite of the intended direction. The optimizer could potentially improve the design by increasing the curvature of these walls, thereby enhancing the contraction mechanism. However, the design-variable bounds and filtering constraints limit this possibility. Attempts to relax these constraints resulted in designs exhibiting self-intersection. Although the operating pressure could be reduced to avoid the ballooning effect, we choose to retain the example in its current form to illustrate this behavior.


\subsection{Inverse design of actuators that support sequential deformation modes} 
Finally, we consider actuators capable of sequential deformation under an applied pressure of $\hat{p}=25$ kPa. Specifically, we design two actuators: one that first grasps an object and subsequently extends in the $e_x$ direction, and another that first grasps an object and subsequently contracts in the $e_x$ direction.
For both optimization problems, we employ two objective functions (i.e., $n_g=2$ in Eq.~\eqref{opt}). For the first objective, we set $\bm{v}^1=(0,-\cos(\theta),-\sin(\theta))$ in Eq.~\eqref{g0} and evaluate it at a load scaling factor of $\lambda^1=0.25$. For the second objective, we set $\bm{v}^2=(1,0,0)$ for axial expansion and $\bm{v}^2=(-1,0,0)$ for axial contraction, and evaluate it at a load scaling factor of $\lambda^2=1.0$.
To promote the desired sequential behavior, we set $\bm{s}^1=(1,0,0)$ and $\varepsilon_u=1$ in the constraint defined by Eq.~\eqref{constraint}. This constraint mitigates deformation in the $e_x$ direction at the first target point, where the actuator is optimized only for the grasping motion. Although the min-max formulation is used within the MMA optimizer, the two objectives must still be weighted differently to achieve the desired behavior. In this example, we set $w_1=-18$ and $w_2=-1$. The filter parameter is set to $G_\psi=3\times10^{-2}$ mm$^{-1}$ in Eq.~\eqref{Wshape}, and the mesh coordinates are updated every 50 design iterations, resulting in a total of four mesh updates.

\begin{figure}
    \centering
    \includegraphics[width=0.9\textwidth]{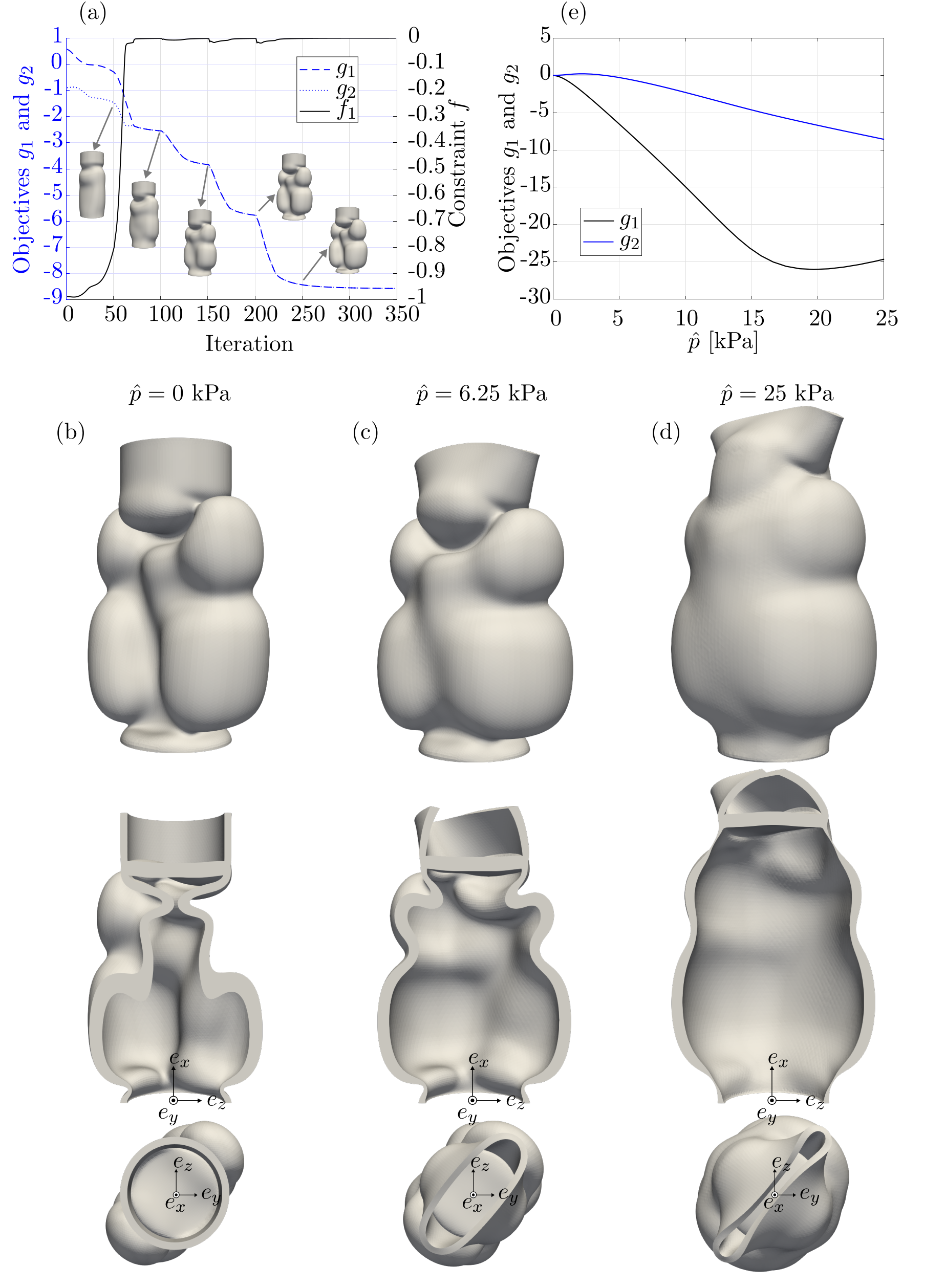}
    \captionof{figure}{\textbf{The gripping and extending actuator}. \textbf{(a)} The evolution of the objective functions $g_1$ and $g_2$, and the constraint $f$ over the design iterations, along with snapshots of intermediate designs. \textbf{(b)} The side-view, sliced-view and top-view of the design in the undeformed configuration, i.e. at $\hat{p}= 0$ kPa. \textbf{(c)} The side-view, sliced-view and top-view of the design at the first target pressure $\hat{p}= 6.25$ kPa. \textbf{(d)} The side-view, sliced-view and top-view of the design at the second target pressure $\hat{p}= 25$ kPa. \textbf{(e)} The objective functions $g_1$ and $g_2$ versus the applied pressure $\hat{p}$.}
    \label{fig:grabExtend}
\end{figure}

Results for the actuator that  that first grasps  and subsequently elongates are shown in Fig.~\ref{fig:grabExtend}. The evolution of the objective functions $g_1$ and $g_2$, as well as the constraint $f$, along with snapshots of the intermediate designs is plotted versus the design iterations in Fig. \ref{fig:grabExtend}(a). The identified optimal design and a sliced view of it is depicted in Fig. \ref{fig:grabExtend}(b) in the undeformed configuration at $\hat{p} = 0$ kPa. The deformed design at the first target pressure $\hat{p} = 6.25$ kPa is depicted in Fig. \ref{fig:grabExtend}(c), and at the second target pressure $\hat{p} = 25$ kPa in Fig. \ref{fig:grabExtend}(d). It is seen that at the first target pressure, the design barely displaces in the $e_x$ directions, as expected by the introduction of the constraint; it focuses on performing the gripping deformation as seen in the top-view in Fig. \ref{fig:grabExtend}(c). At the second target pressure, we instead observe the desired displacement in the positive $e_x$ direction, see Fig. \ref{fig:grabExtend}(d). We curiously note that a larger gripping deformation occurs at the maximum pressure $\hat{p} = 25$ kPa, compared to that of the first, lower, target pressure $\hat{p} = 6.25$ kPa where this deformation mode is actually maximized. Based on this observation, we assume that the actuator design that maximizes the gripping deformation at the maximum pressure, also maximizes this deformation mode at the lower pressure. In Fig. \ref{fig:grabExtend}(e), we plot of the objectives $g_1$ and $g_2$ versus the applied pressure $\hat{p}$. Herein, we confirm the above observations; that the magnitude of $g_2$ is small around $\hat{p} = 6.25$ kPa where we introduce the constraint, and that the magnitude of $g_1$ is larger around $\hat{p} = 25$ kPa than around $\hat{p} = 6.25$ kPa.


\begin{figure}
    \centering
    \includegraphics[width=0.9\textwidth]{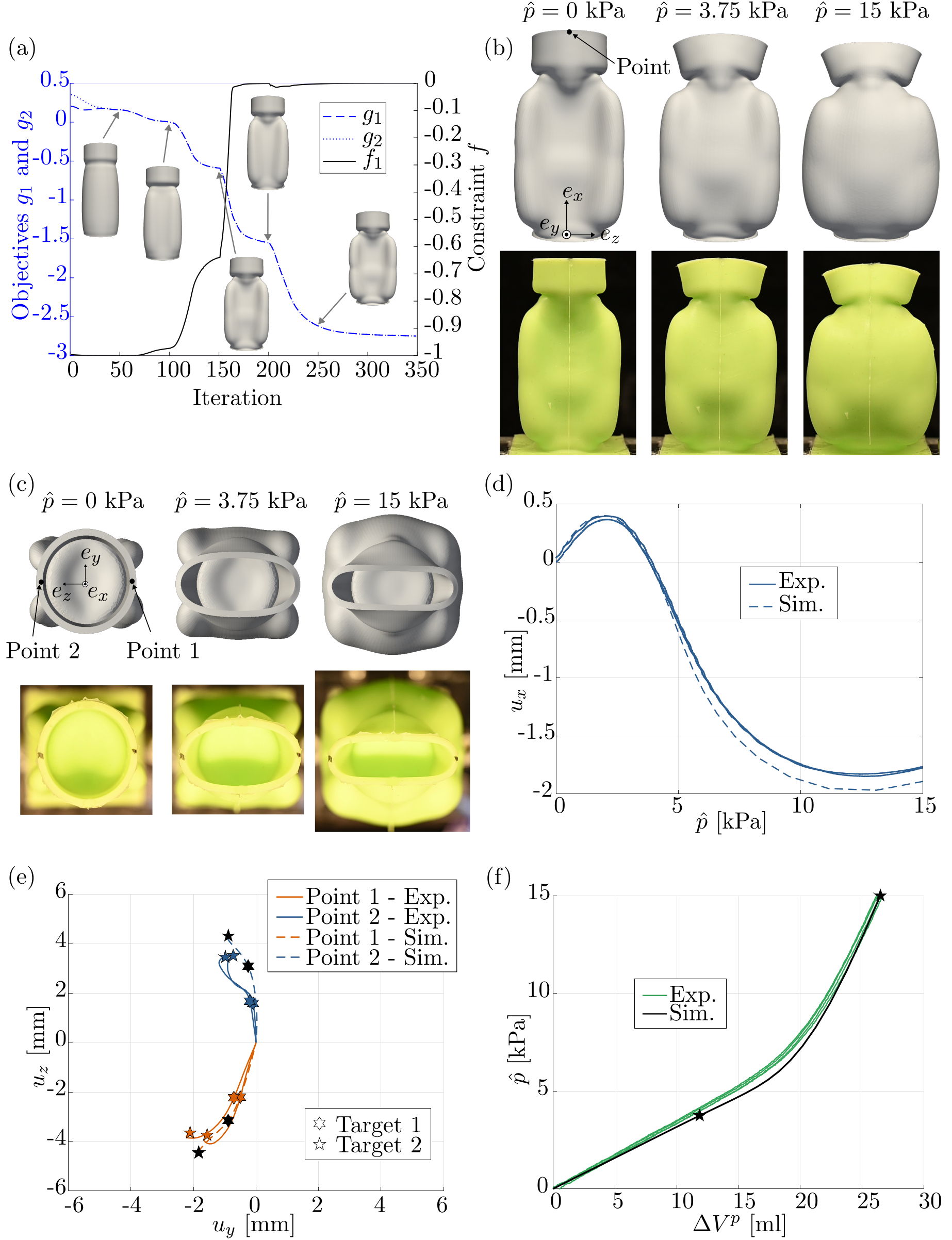}
    \captionof{figure}{\textbf{The gripping and contracting actuator}. \textbf{(a)} The evolution of the objective functions $g_1$ and $g_2$, and the constraint $f$  over the design iterations, along with snapshots of intermediate designs. \textbf{(b)} The side-view of the design and its realization at $\hat{p}= 0$ kPa, $\hat{p}= 3.75$ kPa and $\hat{p}= 15$ kPa. \textsf{(c)} The top-view of the design and its realization at $\hat{p}= 0$ kPa, $\hat{p}= 3.75$ kPa and $\hat{p}= 15$ kPa. \textbf{(d)} The displacement in the $e_x$ direction of the point illustrated in (b) during pressurization. \textbf{(e)} The displacement of the two points illustrated in (c)  in the $e_y$-$e_z$ plane during pressurization. \textbf{(f)} The applied pressure $\hat{p}$ versus cavity volume change $\Delta V^p$.}
    \label{fig:grabContract}
\end{figure}

Finally, in Fig.~\ref{fig:grabContract} we report numerical and experimental results  for an actuator that first grasps  and subsequently contracts.   The evolution of the objective functions $g_1$ and $g_1$, as well as the constraint $f$, is plotted versus the design iterations in Fig. \ref{fig:grabContract}(a), where also snapshots of the intermediate designs are included. A side-view and a top-view of the optimized design and its cast realization are depicted in Fig.~\ref{fig:grabContract}(b) and Fig.~\ref{fig:grabContract}(c) in their undeformed configuration at $\hat{p}=0$~kPa, and deformed configurations at target pressures $\hat{p}=3.75$~kPa and $\hat{p}=15$~kPa. Again, we perform motion tracking of the points highlighted in Fig. \ref{fig:grabContract}(b)-(c) and compare the experimental and simulated displacements over time. The results are depicted in Fig. \ref{fig:grabContract}(d)-(e) and in the movie provided in the supplementary material, where we plot the displacements trajectories during pressurization. A glance at Fig. \ref{fig:grabContract}(b)-(c) confirms that the overall experimental deformation patterns agrees well with the simulation. However, we again note slight discrepancies between the numerical and experimental displacements trajectories in Fig. \ref{fig:grabContract}(d)-(e). To further characterize the actuator, we compare the numerically computed and experimentally measured pressure-volume relation in Fig. \ref{fig:grabContract}(f), which show good agreement. Similarly to the contracting actuator, we observe a non-linear pressure-volume relation. 

\section{Conclusions}
In this work, we design soft actuators using a three-dimensional shape optimization framework. The simulation-based framework is founded on hyperelasticity, nonlinear kinematics, and mixed finite elements, and is implemented in PETSc. The numerical model is shown to agree well with experimental measurements. While minor local deviations are observed in the deformation response, the global behavior exhibits excellent agreement. The optimized designs demonstrate tailored staggered deformations in both simulations and experiments.

Future work will focus on incorporating additional manufacturing constraints into the optimization framework. Demolding has proven to be a critical step in the fabrication process, and curvature constraints may improve manufacturability \citep{schmitt2017curvature}. Furthermore, the robustness of the shape optimization framework could be enhanced by mitigating surface self-intersections through contact-aware optimization techniques \citep{sjovall2025contact}. Another important direction is the incorporation of contact mechanics into the optimization framework to enable the direct optimization of grasping performance. In this context, the third-medium contact method \citep{bluhm2021internal,dahlberg2026rotation} represents a promising approach for modeling contact interactions.

\section{Acknowledgments}
This work was performed under the auspices of the Swedish research council (grant nbr. 2024-00172). The numerical computations were enabled by resources provided by LUNARC, The Centre for Scientific and Technical Computing at Lund University. Last but not least, the authors would like to thank Alex Zhang and Leon Kamp for help with the experimental setup. 

\section*{Appendix}

\subsection*{Evaluation of pressure load}
To evaluate the external pressure load, we introduce the isoparametric mapping $\bm{x} = \bs{N}_\xi(\xi_1, \xi_2)\bs{x}^e$ for all $\bm{x}\in\partial\Omega_c$, where $\bs{N}_\xi$ are the element shape functions corresponding to a surface element, $\bs{x}^e$ are the deformed element coordinates and $\xi_1$ and $\xi_2$ are the isoparametric coordinates. We introduce the deformed current domain $\Omega_c$ with coordinates $\bm{x}\in\Omega_c$ and boundary $\partial\Omega_c$ with normal $\bm{n}_c$, defined as
\begin{equation}
\bm{n}_c = \frac{\pfrac{\bm{x}}{\xi_1}\times\pfrac{\bm{x}}{\xi_2}}{\norm{\pfrac{\bm{x}}{\xi_1}\times\pfrac{\bm{x}}{\xi_2}}_2} = \frac{\hat{\bm{n}}_c}{\norm{\hat{\bm{n}}_c}_2}.
\label{normal}
\end{equation}
The element external load due to the pressure load is therefore
\begin{equation}
\bs{F}_{p} = \fesum\int_{\partial\Omega_c}  p \bs{N}_\xi^T \bm{n}_c \, dS = \fesum\int_{\Omega_\xi}  p \bs{N}_\xi^T \frac{\hat{\bm{n}}_c}{\norm{\hat{\bm{n}}_c}_2} \norm{\hat{\bm{n}}_c}_2 \, d\xi_1 d\xi_2 = \fesum\int_{\Omega_\xi}  p \bs{N}_\xi^T \hat{\bm{n}}_c  \, d\xi_1 d\xi_2,
\label{isoPressureLoad}
\end{equation}
where $\Omega_\xi$ is the isoparametric domain, and we used that $dS = \norm{\hat{\bm{n}}}_2 \, d\xi_1 d\xi_2$. In the above, $p = 0.2$ is the total applied pressure. The stiffness matrix contribution from the pressure load is
\begin{equation}
\bs{K}_p = \fesum \pfrac{{\bs{F}_p}}{\bs{u}^e} =  \fesum \int_{\Omega_\xi}  p \bs{N}_\xi^T \pfrac{\hat{\bm{n}}_c}{\bs{u}^e}  \, d\xi_1 d\xi_2,
\label{dPdshape}
\end{equation}
where the components of $\pfrac{\hat{\bm{n}}_c}{\bs{u}^e}$ are
\begin{equation}
\begin{array}{ll}
\ds\pfrac{\hat{n}_i}{\textsf{u}_p^\beta} &\ds= \varepsilon_{ijk} \left(\pfrac{}{\textsf{u}_p^\beta}\left(\pfrac{x_j}{\xi_1}\right)\pfrac{x_k}{\xi_2} + \pfrac{x_j}{\xi_1}\pfrac{}{\textsf{u}_p^\beta}\left(\pfrac{x_k}{\xi_2}\right) \right) \\[15pt] 
&\ds = \varepsilon_{ijk} \left(\pfrac{N^\alpha}{\xi_1} \pfrac{\textsf{u}_j^\alpha}{\textsf{u}_p^\beta}\pfrac{x_k}{\xi_2} + \pfrac{x_j}{\xi_1}\pfrac{N^\alpha}{\xi_2} \pfrac{\textsf{u}_k^\alpha}{\textsf{u}_p^\beta} \right)\\[15pt] 
&\ds = \varepsilon_{ipj} \pfrac{N^\beta}{\xi_1} \pfrac{x_j}{\xi_2} + \varepsilon_{ijp} \pfrac{x_j}{\xi_1}\pfrac{N^\beta}{\xi_2}\\[15pt] 
&\ds = \varepsilon_{ipj} \left(\pfrac{N^\beta}{\xi_1} \pfrac{x_j}{\xi_2} -  \pfrac{x_j}{\xi_1}\pfrac{N^\beta}{\xi_2}\right).
\end{array}
\label{dnhatdshape}
\end{equation}
and subscript and superscript indices denote dimensional and nodal indices, respectively.

\subsection*{Shape sensitivity of pressure load}
The pressure force sensitivity is
\begin{equation}
\pfrac{\bm{\mu}_a^T\bs{F}_{p}}{\bm{\psi}^e} = \fesum \int_{\Omega_\xi}  p \left(\bm{\mu}_a^e\right)^T\bs{N}_\xi^T \pfrac{\hat{\bm{n}}_c}{\bm{\psi}^e}  \, d\xi_1 d\xi_2,
\label{shapePressureLoad}
\end{equation}
where $\pfrac{\hat{\bm{n}}_c}{\bm{\psi}^e}$ coincides with \eqref{dnhatdshape}. The shape sensitivity of the internal force appears in \citet{dalklint2024simultaneous}.

\subsection*{Shape sensitivity of objective function}
The shape sensitivity of the objective function is 
\begin{equation}
\pfrac{g_i}{\bm{\psi}^e} = -\frac{1}{\vert\partial\Omega^d\vert^2} \pfrac{\vert\partial\Omega^d\vert}{\bm{\psi}^e} \int_{\partial\Omega^d} \bm{u} \cdot \bm{v} \, dS+ \frac{1}{\vert\partial\Omega^d\vert} \int_{\partial\Omega^d} \bm{u} \cdot \pfrac{\bm{v}}{\bm{\psi}^e} \, dS + \frac{1}{\vert\partial\Omega^d\vert} \int_{\partial\Omega^d} \bm{u} \cdot \bm{v} \, \pfrac{dS}{\bm{\psi}^e},
\label{g0sens}
\end{equation}
where $\pfrac{\bm{v}}{\bm{\psi}^e}$ is problem specific and
\begin{equation}
\pfrac{\vert\partial\Omega^d\vert}{\bm{\psi}^e} = \int_{\partial\Omega^d}\, \pfrac{dS}{\bm{\psi}^e}.
\label{dSsens}
\end{equation}
In the above, we evaluate
\begin{equation}
\pfrac{dS}{\bm{\psi}^e} = \pfrac{\norm{\hat{\bm{n}}}_2}{\bm{\psi}^e} \, d\xi_1 d\xi_2 = \frac{1}{\norm{\hat{\bm{n}}}_2} \pfrac{\hat{\bm{n}}}{\bm{\psi}^e} \cdot \hat{\bm{n}} \, d\xi_1 d\xi_2 = \pfrac{\hat{\bm{n}}}{\bm{\psi}^e} \cdot \bm{n} \, d\xi_1 d\xi_2,
\label{dSsens2}
\end{equation}
where the components of $\pfrac{\hat{\bm{n}}}{\bm{\psi}^e}$ are
\begin{equation}
\begin{array}{ll}
\ds\pfrac{\hat{n}_i}{\psi_p^\beta} &\ds= \varepsilon_{ijk} \left(\pfrac{}{\psi_p^\beta}\left(\pfrac{X_j}{\xi_1}\right)\pfrac{X_k}{\xi_2} + \pfrac{X_j}{\xi_1}\pfrac{}{\psi_p^\beta}\left(\pfrac{X_k}{\xi_2}\right) \right) \\[15pt] 
&\ds = \varepsilon_{ijk} \left(\pfrac{N^\alpha}{\xi_1} \pfrac{\psi_j^\alpha}{\psi_p^\beta}\pfrac{X_k}{\xi_2} + \pfrac{X_j}{\xi_1}\pfrac{N^\alpha}{\xi_2} \pfrac{\psi_k^\alpha}{\psi_p^\beta} \right)\\[15pt] 
&\ds = \varepsilon_{ipj} \pfrac{N^\beta}{\xi_1} \pfrac{X_j}{\xi_2} + \varepsilon_{ijp} \pfrac{X_j}{\xi_1}\pfrac{N^\beta}{\xi_2}\\[15pt] 
&\ds = \varepsilon_{ipj} \left(\pfrac{N^\beta}{\xi_1} \pfrac{X_j}{\xi_2} -  \pfrac{X_j}{\xi_1}\pfrac{N^\beta}{\xi_2}\right).
\end{array}
\label{dnhatdshape2}
\end{equation}

\bibliographystyle{elsarticle-harv}
\bibliography{database}

\end{document}